\documentclass[12pt,epsf,epsfig]{article}
\begin{document}

\def\beq{\begin{equation}}
\def\eeq{\end{equation}}
\newcommand{\form}[1]{(\ref{#1})}

\begin{centering}
\begin{flushright}
astro-ph/0309326 \\
CERN-TH/2003-101 \\
April 2003
\end{flushright}

\vspace{0.1in}

{\Large {\bf Development of the Universe and \\ New Cosmology}}

\vspace{0.4in}

{\bf A.S.~Sakharov$^{ab}$ and {\bf H.~Hofer$^a$}}

\vspace{0.2in}

$^a${\em Swiss Institute of Technology, ETH-Zuerich, 8093 Zuerich, Switzerland}\\
$^b${\em TH Division, CERN, 1211 Geneva 23, Switzerland}\\

\vspace{0.4in}
\begin{center} {\bf Abstract} \end{center}
\end{centering}

\vspace{0.2in}

{\small \noindent
 Cosmology is undergoing an explosive period of activity, fueled both by new,
 accurate astrophysical
data and by innovative theoretical developments. Cosmological parameters such as
 the total
density of the Universe and the rate of cosmological expansion are being
 precisely measured for
the first time, and a consistent standard picture of the Universe is beginning
to emerge. Recent developments in cosmology
give rise the intriguing possibility that all structures in the Universe, from
superclusters to planets, had a quantum-mechanical origin in its earliest
moments. Furthermore, these ideas are not idle theorizing, but
 predictive, and subject to meaningful experimental test.
We review the
concordance model of the development of the Universe, as well as evidence for the observational revolution
that this field is going through. This already provides us with important information on particle physics, which is inaccessible to accelerators.}

\vspace{0.8in}
\begin{flushleft}
CERN-TH/2003-101 \\
April 2003
\end{flushleft}

\vspace{0.5in}
\begin{flushleft}


\end{flushleft}


\newpage

\section{Introduction}
Our present understanding of the Universe is based upon the successful
hot Big Bang theory, which explains its evolution from the first
fraction of a second to our present age, around 13 billion years
later. This theory rests upon four strong pillars, a theoretical
framework based on general relativity, as put forward by Albert
Einstein~\cite{Einstein} and Alexander A. Friedmann~\cite{Friedmann} in
the 1920s, and three robust observational facts: First, the expansion of
the Universe, discovered by Edwin P. Hubble~\cite{Hubble} in the 1930s,
as a recession of galaxies at a speed proportional to their distance
from us. Second, the relative abundance of light elements, explained by
George Gamow~\cite{Gamow} in the 1940s, mainly that of helium, deuterium
and lithium, which were cooked from the nuclear reactions that took
place at around a second to a few minutes after the Big Bang, when the
Universe was a few times hotter than the core of the Sun. Third, the
cosmic microwave background (CMB), the afterglow of the Big Bang,
discovered in 1965 by Arno A. Penzias and Robert W. Wilson~\cite{Wilson}
as a very isotropic black-body radiation at a temperature of about 3
degrees Kelvin, emitted when the Universe was cold enough to form
neutral atoms and photons decoupled from matter, approximately 400,000
years after the Big Bang. Today, these observations are confirmed to
within a few per cent accuracy, and have helped establish the paradigm of
 inflationary hot
Big Bang cosmology as the preferred model of the Universe \cite{Bridle:2003yz}.

\section{The Universe as we see it}
All of modern cosmology stems essentially from an application of the Copernican
 principle:
we are not at the centre of the Universe. In fact, today we take Copernicus'
 idea one
step further and assert the ``cosmological principle'': {\em nobody} is at the
 centre of
the Universe. The cosmos, viewed from any point, looks the same as when viewed
 from any
other point. This, like other symmetry principles more directly familiar to
 particle
physicists, turns out to be an immensely powerful idea. In particular, it leads
 to the
apparently inescapable conclusion that the Universe has a finite age. There was
 a
beginning of time.

This cosmological principle can be expressed by applying one of the most
``reasonable'' symmetries to the Einstein equation, which initially connects the
geometry of space-time with the matter content of the Universe. The simplest
symmetries of this type are {\it homogeneity } and {\it isotropy}. By
homogeneity, we mean that the Universe is invariant under spatial translations,
while  isotropy implies invariance under rotations. In this sense the contents of
the Universe can be modelled as a perfect fluid with some equation of state
\footnote{The equation of state is frequently given by a relation between
density $\rho$ and pressure $p$ of the fluid.}. While this is certainly a poor
 description of
the contents of the Universe on small scales, such as the size of people or
planets or even galaxies, it is an excellent approximation if we average over
extremely large scales in the Universe, for which the matter is known
observationally to be very smoothly distributed.

General relativity combined with homogeneity and isotropy leads to a startling
conclusion: space-time is dynamic. The Universe is not static, but is bound to
be either expanding or contracting. Indeed, in 1929, Hubble undertook a
project to measure the distances to the spiral ``nebulae'', as they had been
known. Hubble's method involved using Cepheid variables. Cepheid variables have
the useful property that the period of their variation is correlated to their
absolute brightness. Therefore, by measuring the apparent brightness and the
period of a distant Cepheid, one can determine its absolute brightness and
therefore its distance. Hubble applied this method to a number of nearby
galaxies, and determined that almost all of them were receding from the Earth.
Moreover, the more distant the galaxy was, the faster it was receding, according
to a roughly linear relation: $ v = H_0 d. $
This is the famous Hubble law, and the constant $H_0$ is known as Hubble's
constant.  The current best estimate, determined using the Hubble space
telescope to resolve Cepheids in galaxies at unprecedented distances, is $H_0 =
71 \pm 6\ {\rm km\cdot s^{-1}\cdot Mpc^{-1}}$ \footnote{The {\em parsec} is a spatial  astronomical
unit corresponding to one second of arc of the parallax measured from opposite sides
of the earth's orbit: $1\ {\rm pc} =3\cdot 10^{18}{\rm cm}$.}
\cite{Freedman99}. In any case, the Hubble law is exactly what one would expect
from the so-called expanding Friedmann--Robertson--Walker (FRW) Universe
\cite{Weinberg}.

The expansion of the Universe leads to a number of interesting effects. One is
 the
cosmological redshift of photons. The usual way to see this is that, from the
 Hubble law,
distant objects appear to be receding at a velocity $v = H_0 d$, which means
 that photons
emitted from the body are redshifted by the recession velocity of the
 source.  There is
another way to look at the same effect: because of the expansion of space, the
 wavelength
of a photon increases with the scale factor:
$
\lambda \propto a(t),
$
so that a photon, propagating in the space as the universe expands,  gets shifted
 to longer
and longer wavelengths. The redshift $z$ of a photon is then given by the ratio
 of the scale
factor today to the scale factor when the photon was emitted:
$
1 + z = {a(t_0) \over a(t_{\rm em})}.
$
This redshifting due to expansion applies
to particles other than photons as well. For some massive body moving relative
 to the
expansion with some momentum $p$, the momentum also ``redshifts'':
$
p \propto {1 \over a(t)}.
$
We then have the remarkable result that freely moving bodies in an expanding
 Universe eventually
come to rest relative to the expanding coordinate system.
Thus the expansion of the Universe creates a kind of dynamical friction for
 everything moving
in it. If we take a bunch of particles  with temperature $T$ in thermal
equilibrium, the momenta of all these particles will
 decrease linearly
with the expansion, and the system will cool. For a gas in
thermal equilibrium, the temperature is in fact inversely proportional to the
 scale factor:
$
T \propto {1 \over a(t)}.
$
The current temperature of the Universe is $2.725\pm 0.005\ {\rm K}$
 \cite{Spergel:2003cb}.

One of the things that cosmologists most want to measure accurately is the total
 density
$\rho$ of the Universe. This is very often expressed in units of the
critical density needed to make the geometry of the Universe flat.  Observers
have made attempts to measure the density of the Universe using a variety of
methods, including measuring galactic rotation curves, the velocities of
galaxies orbiting in clusters, X-ray measurements of galaxy clusters, the
velocities and spatial distribution of galaxies on large scales, and
gravitational lensing. These measurements have repeatedly pointed to the
existence of a large amount of dark matter, different from the baryonic
constituents of stars and planets. Moreover two groupes - the Supernova Cosmology Project
\cite{SCPHubblediag} and the High-z Supernova Search Team \cite{hz} has recently arrived  at a picture of the Universe with
 flat geometry: matter,
including both baryons and the mysterious dark matter,
makes up only about 30\% of the energy density in the Universe. The remaining
 70\% is
made of something that looks very much like a fluid with negative
pressure, which  slightly accelerate the current expansion of the Universe \footnote{Deeper and more precies supernova measurements \cite{tonry} confirm this conclusion. Future data on supernova from space based \cite{hstsuper,snap} and ground based \cite{ground} telescopes promise to yield substantial
information on the history of cosmic acceleration, and to constrain possible models for
the dark energy.}.
 This {\em dark energy} can possibly be identified with the vacuum
energy predicted by quantum field theory, except that the energy density is 120
 orders
of magnitude smaller than would be expected from a naive analysis. This density
budget of the Universe, where the critical density is distributed between the
 following contributions: $4.7\pm 0.6$\% luminous matter, $24\pm 7$\% dark
 matter
and $71.3\pm 8$\% of dark energy, gets confirmed at very high confidence
 level by
recent fascinating results of the WMAP satellite, which in fact "took a
picture" of the Universe when it was only about 400,000 years old.

\section{Thermal history of the Universe and beyond}
The basic picture of an expanding, cooling Universe leads to a number of
startling predictions: the formation of nuclei with its resulting primordial
abundances of elements, and the later formation of neutral atoms with the
consequent presence of a cosmic background of photons \cite{CMBreview,CMBbib}.

 As we go back in time, the
Universe becomes hotter and hotter and thus the amount of energy
available for particle interactions thus increases. As a consequence, the
nature of interactions goes from those described at low energy by long-range gravitational and electromagnetic physics, to atomic physics,
nuclear physics, all the way to high energy physics at the electroweak
scale, gran unification (perhaps), and finally quantum gravity.

The way we know about the high energy interactions of matter is via
particle accelerators, which are unravelling the details of these
fundamental interactions as we increase their energy. However, one should
bear in mind that the physical conditions that take place in our high
energy colliders are very different from those that occurred in the
early Universe: these machines could never reproduce the conditions of
density and pressure in its rapidly expanding thermal plasma. Nevertheless, those experiments are crucial in
understanding the nature and {\em rate} of the local fundamental
interactions available at those energies. What interests cosmologists is
the statistical and thermal properties that such a plasma should have,
in particular the time at
which certain particles {\em decoupled} from the plasma, i.e. when their
interactions were not quick enough with respect to the expansion of the
Universe, and they were left out of equilibrium with the plasma.

One can trace the evolution of the universe from its origin till today.
 According to the best accepted view, the
Universe must have originated at the Planck era ($10^{19}$ GeV,
$10^{-43}$ s), from a quantum gravity fluctuation. Needless to say, quantum
gravity phenomena are still in the realm of physical speculation, although the last
astrophysical probes can  already feel effects of quantum gravity at the energy
 level of the Grand Unified Theories (GUTs), about $10^{16}$~GeV
($10^{-35}$ s) \cite{we} and beyond \cite{Jacobson:2002ye}. However, it is plausible that a pre Big Bang era of
cosmological {\em inflation} originated then. The idea of inflation, a period of
accelerated expansion before the FRW expansion is established, provides an
elegant solution to set up the initial conditions for standard Big Bang
cosmology \cite{linde}. At some early time, just before the Universe may have
reached the GUTs era ($10^{16}$ GeV, $10^{-35}$ s) and
was not yet thermalized, the energy density of the Universe was dominated by some
material ({\em inflaton field}) with negative pressure. Under such an assumption, the
Einstein equation leads to the conclusion that during the pre Big Bang era the
expansion of the Universe was exponentially accelerated, spreading out the
conditions that took place in a small causally connected region over a huge
distance across which we observe the Universe now. The last concequence explains, in
 particular, how became the
Universe so big and so uniform before becoming the FRW Universe.

Quantum fluctuations of the
inflaton field then left their imprint as tiny
perturbations in an otherwise very homogeneous patch of the Universe. One of the
most astonishing predictions of inflation is that quantum fluctuations of
the inflaton field, being stretched by the exponential expansion generate
large-scale perturbations of the space-time geometry. Patterns of
perturbations in the geometry are like fingerprints that unequivocally
characterize a period of inflation. When matter falls in the gravitational weels
of the patterns, it creates density perturbations that collapse gravitationally
to form galaxies, clusters and superclusters of galaxies. Perhaps the most
intriguing aspect of modern cosmology is the fact that the large-scale structure
was seeded at the extremely early epoch of the development of the Universe,
when even the conception of matter in usual sense did not exist yet. Moreover
this fact is already tested observationally at a very high level of accuracy by
the combination of recent results from WMAP satellite \cite{Spergel:2003cb} andthe 2dF
galaxy redshift survey \cite{Peacock:2003ah}.

At
the end of inflation, the huge energy density of the inflaton field was
converted into particles, which soon thermalized and became the origin
of the hot Big Bang. Such a process is called {\em
reheating} of the Universe. Since then, the Universe became 
radiation-dominated.

It is probable (although by no means certain) that the
asymmetry between matter and antimatter originated at the same time as
the rest of the energy of the Universe, from the decay of the
inflaton. This process is known under the name of {\em baryogenesis} \cite{Dine:2003ax},
since baryons (mostly quarks at that time) must have originated then,
from the leftovers of their annihilation with antibaryons.
Any mechanism of baryogenesis requires a violation of the baryon number, C and CP
violation, and a departure from thermal equilibrium \cite{Sakharov:dj}. The
first two conditions can be discussed only within a particle physics model that
 is
definitely beyond the Standard Model. This fact represents perhaps the best
example of the perfect interplay between cosmology and particle physics.

 As the Universe cooled down, it may have gone through the
quark--gluon phase transition ($10^2$ MeV, $10^{-5}$ s), when baryons
(mainly protons and neutrons) formed from their constituent quarks \cite{Schwarz:2003du}. A minor
contribution of antimatter regions, which might be left from baryogenesis, can
evolve into condensed antimatter objects; these are important footprints of early
phase transitions, which took place far above electroveak energies
\cite{Khlopov:2002ww}. In this sense the search for antimatter in space, with
the AMS-02 experiment \cite{ams}, turns out to be an important step into a deeper
 understanding of
physics beyond the Standard Model which took place in the early Universe.

The furthest window we have on the early Universe at  is for the moment that
of {\em primordial nucleosynthesis} ($1$--$0.1$ MeV, 1 s -- 3 min), when
protons and neutrons were cold enough that bound systems could form,
giving rise to the lightest elements, soon after {\em neutrino
decoupling}: It is the realm of nuclear physics. The observed relative
abundances of light elements are in agreement with the predictions of
the hot Big Bang theory. Immediately afterwards, electron--positron
annihilation occurs (0.5 MeV, 1 min) and all their energy goes into
photons. Much later, at about 1 eV, $\sim 10^5$ yr, matter and
radiation have equal energy densities. Soon after, electrons become
bound to nuclei to form atoms (0.3 eV, $3\times10^5$ yr), in a process
known as {\em recombination}: this is the realm of atomic physics.

Immediately after, photons decouple from the plasma, travelling freely
thereafter. Those are the photons we observe as the CMB. The COBE satellite
 \cite{cobe} launched in 1990 observed a tiny anisotropy in the angular
 distribution of the CMB temperature. It is believed that
this anisotropy represents intrinsic fluctuations in the CMB itself, due to the
presence of tiny primordial density fluctuations in the cosmological matter
present at the time of recombination. These density fluctuations are of great
physical interest, because these are the fluctuations that later
collapsed to form all of the structures in the Universe, from superclusters to
planets. Moreover the angular
distribution of the CMB depends, for example, on the baryon
density, the Hubble constant $H_0$, the densities of dark matter
 and dark energy. This
makes the interpretation of the angular spectrum something of a complex undertaking,
 but it also makes it a sensitive probe of cosmological
models and, implicitly, a particle physics laboratory of high precision
\cite{ellis}.

Much later ($\sim 1$--$10$
Gyr), the small inhomogeneities generated during inflation have grown, via
gravitational collapse, to become galaxies, clusters of galaxies, and
superclusters, characterizing the epoch of {\em structure formation}. It is the
realm of long--range gravitational physics, dominated by a dark (vacuum) energy. Finally (3~K, 13~Gyr), the Sun, the
Earth, and biological life originated from previous generations of stars, and
from a primordial soup of organic compounds, respectively.

The recent announcement by the WMAP satellite team of the landmark measurements of
 the CMB anisotropy \cite{Spergel:2003cb} has convincingly confirmed important
 aspects of the current standard cosmological model described above.

\section{Conclusions}
We have entered a new era in cosmology and astrophysics, where a host of
high-precision measurements are already posing challenges to our understanding
of the Universe: the density of ordinary matter and the total amount of energy
in the Universe; the microwave background anisotropies on a fine-scale
resolution; primordial deuterium abundance from quasar absorption lines;
the acceleration parameter of the Universe from high-redshift supernovae
observations; the rate of expansion from gravitational lensing~\cite{Bartelmann:1999yn}; large-scale structure measurements of the distribution of galaxies and their
evolution; and many more, which already put constraints on the parameter
space of cosmological models.

However, these are
only the forerunners of the precision era in cosmology that will
dominate the new millennium, and make cosmology a phenomenological
science, whose results can be incorporated into experimental high
energy physics \cite{Khlopov:rs}. An impressive example of such a symbiosis of modern observational
 cosmology and experimental particle physics was the last improvement
of the probe of the neutrino masses by 2dF galaxy redshift surveys data combined
with WMAP results. The obtained limit \cite{Spergel:2003cb} on the sum ot the neutrino
masses of 0.7~eV is substantially better than even the most stringent direct
laboratory limit on any individual neutrino mass. On top of it, WMAP data also
provide a new limit on the effective number of light neutrino species, \ beyond
the \ three \ within  the Standard \  Model \cite{lesg}:  $-1.5<\Delta
N_{\nu}^{eff}<4.2$. This limit is not as stringent as that from LEP, but it applies
to additional light particles, that might not be produced in Z decay. The
cosmological upper limit on the masses of supersymmetric particles which can
 play the role of dark matter, is also
improved by WMAP \cite{loh}.

One of the most difficult challenges that the new cosmology will have to
face is understanding the origin of the dark energy, whose behaviour is similar to
 that of the inflaton field, but 13 Gyr later \cite{Saini:1999ba}. It is of
 tremendous
interest from the standpoint of fundamental theory \cite{Padmanabhan:2002ji}. In this sense cosmology provides us with a
 way
to study a question of central importance for particle theory, namely the nature
of the vacuum in quantum field theory. This is something that cannot be studied
in particle accelerators, so in this sense cosmology provides a unique window on
particle physics.

\end{document}